\documentclass[12pt]{article}
\usepackage{graphicx}
\begin{document}

\begin{center}

{\Large \bf Diagonalization of Sp(2) matrices}

\vspace{7mm}

S. Ba{\c s}kal\footnote{electronic address:
baskal@newton.physics.metu.edu.tr}, \\
Department of Physics, Middle East Technical University,
06531 Ankara, Turkey

\vspace{1.ex}
Elena Georgieva\footnote{electronic address: egeorgie@pop500.gsfc.nasa.gov} \\
Goddard Earth Sciences and Technology Center,
University of Maryland Baltimore County,
Baltimore, Maryland 21228, U.S.A.

\vspace{1.0ex}

Y. S. Kim\footnote{electronic address: yskim@physics.umd.edu} \\
Department of Physics, University of Maryland,
College Park, Maryland 20742, U.S.A.

\end{center}

\vspace{2.0ex}

\begin{abstract}
The two-by-two $Sp(2)$ matrix has three parameters with unit
determinant.  Yet, there are no established procedures for
diagonalizing this matrix.  It is shown that this matrix can
be written as a similarity transformation of the two-by-two Wigner
matrix, derivable from Wigner's little group which dictates the
internal space-time symmetries of relativistic particles.  The
Wigner matrix can be diagonalized for massive and space-like
particles, while it takes a triangular form with unit diagonal
elements for light-like particles.  The most immediate physical
application can be made to repeated one-dimensional transfer
matrices appearing in many different branches of physics.  Another
application of current interest could be the dis-entanglement of
entangled systems.
\end{abstract}

\newpage
\section{Introduction}
The two-by-two matrices of the group $Sp(2)$ or its unitary
equivalent appear in almost all branches of physics.  Indeed, a
systematic approach to the study of those matrices is a respectable
branch of physics~\cite{guill84,lang85}.

The $Sp(2)$ matrix is the basic language for linear canonical
transformations for phase-space approach to both classical and
quantum mechanics~\cite{knp91}.  The same is true for classical
and quantum optics~\cite{knp91}, as well as the current
problem of entangled harmonic oscillators~\cite{giedke03,kn05}.
The one-dimensional scattering matrix also takes the form of this
$Sp(2)$ matrix.  The two-by-two beam transfer matrix is unitarily
equivalent to that of the $Sp(2)$ group~\cite{sprung93,monzon00,griff01}.
The list could be endless.  In short, we cannot do modern physics
without two-by-two matrices, especially without those of the $Sp(2)$
group.

Do we then know how to diagonalize this simple mathematical
expression?  We can if they can be diagonalized by a rotation.
If not, there are no standard procedures for approaching this
problem.  In this paper, we address this fundamental issue.  Our
main interest is whether this matrix can be written as a similarity
transformation of a diagonal matrix.  If that is the case, we solve
the practical problem of calculating repeated applications of the
same matrix as we see in finite one-dimensional crystals, including
periodic potentials~\cite{sprung93,griff01}, multilayer
optics~\cite{gk03,sanch05}, lens optics~\cite{bk01,bk03},
and laser cavities~\cite{bk02}, and many other problems in physics.

We first show that the most general form of the $Sp(2)$ matrix
can be written as a similarity transformation of Wigner's
little-group matrix~\cite{wig39}.  This Wigner matrix takes three
different forms depending on the parameters of the original $Sp(2)$
matrix.  However, one of these three matrices is not diagonalizable,
but offers the same conveniences as diagonal matrices do.  Indeed,
the diagonalization of the $Sp(2)$ matrix requires the construction
of a two-by-two representation of Wigner's little groups.

For this purpose, we start with the Bargmann decomposition of
the $Sp(2)$ matrix in which it is written as a product of three
one-parameter matrices.  We also write the matrix as a similarity
transformation of Wigner's little group matrix.  In this way,
the Wigner parameters can be written as the Bargmann parameters,
allowing us to write the $Sp(2)$ matrix as a similarity transformation
of the Wigner matrix, namely Wigner's little-group matrix.

In section~\ref{barg}, we start with the Bargmann representation and
transform it into a form convenient for writing it as a similarity
transformation of the Wigner's little-group matrix discussed in
section~\ref{wig}.  In section~\ref{sp2}, the Bargmann parameters are
written in terms of the Wigner parameters.  The $Sp(2)$ matrix
takes many different forms.  They are shown to be unitarily
equivalent in section~\ref{detail}.  In section~\ref{appl}, we discuss
some immediate applications of the diagonalization by Wignerization
of the $Sp(2)$ matrix.

\section{Bargmann decomposition}\label{barg}
The $Sp(2)$ matrix appears in the physics literature in several
different forms.  This two-by-two unimodular matrix is often
written as
\begin{equation}\label{abcd}
M = \pmatrix{A & B \cr C & D} ,
\end{equation}
with $(AD - BC) = 1.$  All four elements are real numbers.  This
matrix is commonly called the $ABCD$ matrix and has three independent
parameters.

Furthermore, this unimodular matrix $M$ can also be written as
\begin{equation}\label{barg11}
M = R_{1}(\theta_1) B(2\lambda) R_{2}(\theta_2) ,
\end{equation}
with
\begin{eqnarray}\label{barg22}
&{}& B(2\lambda) = \pmatrix{\cosh\lambda & \sinh\lambda \cr
        \sinh\lambda & \cosh\lambda} ,   \nonumber \\[2ex]
&{}& R_{i}(\theta_i) = \pmatrix{\cos(\theta_i/2) & -\sin(\theta_i/2)
\cr   \sin(\theta_i/2)  &  \cos(\theta_i/2)} .
\end{eqnarray}
This is known as the Bargmann decomposition~\cite{barg47}.  We use the
notation $B(2\lambda)$ instead of $B(\lambda)$ purely for convenience.
The main point is that the original three-parameter matrix is decomposed
into three one-parameter matrices.

We can rewrite the Bargmann decomposition of equation~(\ref{barg11})
as
\begin{equation}\label{barg33}
M = R_1 B R_2 = \left(L R \right) B \left(R L^{-1}\right)
              = L (R B R) L^{-1} ,
\end{equation}
where
\begin{equation}
R_1 = LR, \qquad R_2 = RL^{-1},
\end{equation}
and
\begin{eqnarray}\label{r22}
&{}& R(\theta) = \sqrt{R_1 R_2} = \sqrt{R_2 R_1}
                = \pmatrix{\cos(\theta/2)  &
     -\sin(\theta/2) \cr \sin(\theta/2)  &  \cos(\theta/2)} ,
         \nonumber \\[2ex]
&{}& L(\delta) = \sqrt{R_1 \, R_2\,^{-1}} =
  \pmatrix{\cos(\delta/2)  &
  -\sin(\delta/2) \cr \sin(\delta/2)  & \cos(\delta/2)} ,
\end{eqnarray}
with
$$
\theta = \frac{\theta_1 + \theta_2}{2} , \qquad
\delta = \frac{\theta_1 - \theta_2}{2}.
$$
Then, the matrix $M$ can be written as a similarity transformation of
$(R B R)$ with respect to $L$, or a rotation of the system by $L$.
Thus, for all practical purposes, it is sufficient to study the
core matrix $RBR$ with two parameters.  This is still a form of
the Bargmann decomposition.

Now the question is how to diagonalize the core matrix $(RBR)$ and
write it as a similarity transformation.  This core matrix takes the
form
\begin{equation}\label{rbr}
R(\theta)B(2\lambda) R(\theta) = \pmatrix{(\cosh\lambda)\cos\theta &
       - (\cosh\lambda)\sin\theta + \sinh\lambda
      \cr
 (\cosh\lambda)\sin\theta + \sinh\lambda & (\cosh\lambda)\cos\theta} .
\end{equation}
We assume here that both $\sin\theta$ and $\sinh\lambda$ are positive.

The product of the two off-diagonal components becomes
\begin{equation}\label{offd}
\cosh^2\lambda[\tanh\lambda + \sin\theta] [\tanh\lambda - \sin\theta]
= (\cosh\lambda)^2 \cos^2 \theta- 1 ,
\end{equation}
Thus, if one of the off-diagonal components vanishes, the diagonal
elements become one, and the $RBR$ matrix becomes
\begin{equation}\label{iwa11}
H_{+}(\sinh\lambda)=\pmatrix{1  & 0 \cr 2\sinh\lambda & 1} ,
\quad \mbox{or} \quad
H_{-}(\sinh\lambda)=\pmatrix{1  &  -2\sinh\lambda \cr 0 & 1},
\end{equation}
respectively.
This form is known as the Iwasawa decomposition, and is a special case
of the Bargmann decomposition.  This matrix is not diagonalizable.

Yet, we are interested in diagonalizing the $RBR$ matrix by
following the usual procedure of calculating the eigenvalues
of the matrix, which leads to the eigenvalues
\begin{equation}
E_{\pm} = (\cosh\lambda) \cos\theta \pm
\sqrt{(\cosh\lambda)^2\cos^2\theta - 1} .
\end{equation}
with $E_{+} E_{-} = 1 $.  The quantity inside the square root
sign is the same as the product of the two off-diagonal elements
given in equation~(\ref{offd}).  Then the diagonal matrix takes
the form
\begin{equation} \label{dmat}
D = \pmatrix{ E_{+}  & 0 \cr 0 & E_{-}} .
\end{equation}

If $(\cosh\lambda)^2 \cos^2\theta$ is smaller than one, and the
quantity inside the square-root sign is negative,  the eigenvalues
are
\begin{equation}
   E_{\pm} = \exp{(\pm i\phi/2)} ,
\end{equation}
with
\begin{equation}
    \cos(\phi/2) = (\cosh\lambda) \cos\theta .
\end{equation}
If $(\cosh\lambda)^2 \cos^2\theta$ is greater than one, the
eigenvalues become
\begin{equation}\label{dmat2}
   E_{\pm} = \exp{(\pm\chi/2)} ,
\end{equation}
with
\begin{equation}
    \cosh(\chi/2) =  (\cosh\lambda) \cos\theta  .
\end{equation}

If $(\cosh\lambda) \cos\theta$ = 1, the quantity inside the
square-root sign vanishes, the eigenvalues collapse to one, and
the $RBR$ matrix becomes one of the triangular matrices of
equation~(\ref{iwa11}).

In view of the triangular matrices of equation~(\ref{iwa11}), the
best we can do is to write the $RBR$ matrix as a similarity
transformation of
\begin{equation}
\pmatrix{e^{i\phi/2} & 0 \cr  0 &  e^{-i\phi/2} }, \qquad
\pmatrix{e^{\chi/2} & 0 \cr 0 & e^{-\chi/2} },
\end{equation}
or one of the triangular matrices of equation~(\ref{iwa11}).
We shall call the set of these three matrices the Wigner matrix,
use the notation $W_{\pm}$ for the set of three matrices with
$H_{\pm}$, and $W$ collectively for both.

Our plan is two write $RBR$ as a similarity transformation of
the Wigner matrix:
\begin{equation}
RBR = S^{-1} W S .
\end{equation}
As we shall see in Sec.~\ref{wig}, the $W$ matrices are also
derivable from Wigner's little group which dictate the internal
space-time symmetries of elementary particles~\cite{wig39}

Then the question is which among the two Wigner matrices
is to be chosen.  In order to address this question, let us
note the Lorentz group
has two branches.  The $Sp(2)$ group is generated by
\begin{equation}
J_{2} = \frac{1}{2}\pmatrix{0 & -i \cr i & 0} , \quad
K_{3} = \frac{1}{2}\pmatrix{i & 0 \cr 0 & -i} ,  \quad
K_{1} = \frac{1}{2}\pmatrix{0 & i \cr i & 0}  ,
\end{equation}
forming a closed set of commutation relations:
\begin{equation}
\left[J_{2}, K_{3} \right] = iK_{1} ,   \quad
\left[J_{2}, K_{1} \right] = -iK_{2} ,   \quad
\left[J_{2}, K_{3} \right] = -iJ_{2} .
\end{equation}
The rotation generator $J_{2}$ is antisymmetric, and $K_{3}$ and
$K_{1}$ are symmetric.  Furthermore, this set of commutation
relations is invariant under the sign change of the $K_{i}$
matrices while it is not when $J_{2}$ changes its sign.  The
rotation matrices generated by $J_{2}$ matrix are anti-symmetric,
while the squeeze matrices generated by $K_{i}$ are symmetric.

Let us go back to the $RBR$ matrix of equation~(\ref{rbr}).
Since the $B(2\lambda)$ matrix is symmetric and is generated by
$K_{1}$.  We should therefore consider the case where $B(2\lambda)$
is replaced by $B(-2\lambda)$.  The resulting $RBR$ matrix is
\begin{equation}\label{rbrm}
R(\theta)B(-2\lambda) R(\theta) = \pmatrix{(\cosh\lambda)\cos\theta &
      - (\cosh\lambda)\sin\theta - \sinh\lambda \cr
 (\cosh\lambda)\sin\theta - \sinh\lambda & (\cosh\lambda)\cos\theta} .
\end{equation}
For convenience, we shall use the notations $(RBR)_{+}$ and
$(RBR)_{-}$ for equations~(\ref{rbr}) and (\ref{rbrm}) respectively:
\begin{eqnarray}
&{}& (RBR)_{+} = R(\theta) B(2\lambda) R(\theta), \nonumber \\[1ex]
&{}& (RBR)_{-} = R(\theta) B(-2\lambda) R(\theta) .
\end{eqnarray}
We shall still use the notation $RBR$ collectively for $(RBR)_{+}$
and $(RBR)_{-}$.  We cannot obtain $(RBR)_{-}$ from $(RBR)_{+}$ 
by simply changing the sign of the $\lambda$ parameter.  
It is a parity operation.

If both $\sin\theta$ and $\sinh\lambda$ are positive, the upper-right
element of $(RBR)_{+}$ can vanish, while the lower-left element can
become zero for $(RBR)_{-}$.  They are therefore consistent with
$H_{+}(\sinh\lambda)$ and $H_{-}(\sinh\lambda) $ respectively.

Let us see why the $W$ matrix should be called the Wigner matrix
in section~\ref{wig}.

\section{Wigner's little groups}\label{wig}

Since the group $Sp(2)$ is locally isomorphic to the Lorentz
group applicable to two space dimensions and one time variable,
we can import useful matrix identities from the kinematics of
the Lorentz group, particularly from Wigner's little
group~\cite{wig39}.

Wigner's little group is the maximal subgroup of the Lorentz
group whose transformations leave the given four-momentum of
a particle invariant~\cite{wig39}.  Let us first consider a
particle moving along the negative $z$ direction with the
four-momentum
\begin{equation}\label{vec11}
  P = (E, 0, 0, -p) ,
\end{equation}
with $E = \sqrt{p^2 + m^2}$.  We use here the four-vector
convention $(t, x, y, z)$, and the $c = \hbar = 1$ unit system.

It is possible to bring this four-momentum to
\begin{equation}\label{vec22}
  (m, 0, 0, 0) ,
\end{equation}
by applying the transformation matrix
\begin{equation}\label{s44}
S(\eta) = \pmatrix{\cosh\eta  & 0 & 0 &  \sinh\eta \cr
            0  & 1 & 0 &  0 \cr
            0  & 0 & 1 & 0\cr
            \sinh\eta  & 0 & 0 & \cosh\eta } ,
\end{equation}
width
\begin{equation}
E = m (\cosh\eta), \quad  p = m (\sinh\eta) .
\end{equation}

The four-momentum of equation~(\ref{vec22}) is invariant
under the rotation
\begin{equation}\label{rfi}
R(\phi) = \pmatrix{1  & 0 & 0 &  0 \cr
            0  & \cos\phi & 0 &  \sin\phi \cr
               0  & 0 & 1 & 0\cr
            0  & -\sin\phi & 0 &  \cos\phi }  .
\end{equation}

After this rotation, we can bring back the four momentum
of equation~(\ref{vec22}) to the starting momentum
of equation~(\ref{vec11}) by applying the inverse of
the boost matrix $S(\eta)$.

Thus the transformation $S(-\eta) R(\phi) S(\eta)$ will leave
the four-momentum of equation~(\ref{vec11}) invariant.

We can also rotate the entire system around the $z$ axis without
changing the four-momenta of equation~(\ref{vec11}) and
equation~(\ref{vec22}), and the matrix $S(\eta)$.  Thus, we can
obtain the three-dimensional rotation group by taking into account
this degree of freedom along with the rotation matrix of
equation~(\ref{rfi})~\cite{hks86jm}.  The group represented by
this Lorentz-boosted rotation matrix is known as Wigner's
$O(3)$-like little group for massive particles.

If the four-momentum is space-like, it can be brought to
$(0,~0,~0,~p)$.  This four-momentum is invariant under the boost
\begin{equation}\label{xchi}
X(\chi) = \pmatrix{\cosh\chi & \sinh\chi  & 0 & 0 \cr
                    \sinh\chi & \cosh\chi  & 0 & 0 \cr
                    0 & 0   & 1 & 0 \cr 0 & 0  & 0 & 1 } ,
\end{equation}
along the $x$ direction.
The little group is thus represented by $S^{-1} X(\chi)S,$
along with rotations around the $z$ axis.  Wigner's little group
for space-like four-momentum is a Lorentz-boosted $O(2,1)$ group.

If the particle is massless, there are no Lorentz frames in which
the particle is at rest, but its four-momentum can be brought to
$(k,~0,~0,~k)$.  This four-vector is invariant under rotations
around the $z$ axis.  In addition, it is invariant under
\begin{equation}\label{ngamma}
N(\gamma) = \pmatrix{1 + \gamma^2/2 & - \gamma & 0 & -\gamma^2/2 \cr
          - \gamma &  1 & 0 & \gamma \cr 0 & 0 &  1 & 0 \cr
            \gamma^2/2 & \gamma &  0 &   1 - \gamma^2/2} .
\end{equation}
In his 1939 paper, Wigner showed that this $N$ operator together
with the above-mentioned rotation around the $z$ axis form
a three-parameter group which is locally isomorphic to the 
two-dimensional Euclidean group.  Its rotational degree of freedom 
corresponds to the helicity of the particle while the two-translational 
degrees collapse into one gauge degree of freedom~\cite{kiwi90jm}.

This summarizes what Wigner did in his 1939 paper and later papers
on massless particles~\cite{wig39,kiwi90jm}.  The transformation
matrix which leaves the four-momentum invariant is
\begin{equation}\label{wlg11}
S(-\eta) W S(\eta) ,
\end{equation}
where $W$ takes the form of $R(\phi)$, $X(\chi)$, or $N(\gamma)$.

It is well-known that the six-parameter Lorentz group can also be
represented by two-by-two unimodular matrices.  This group of
two-by-two matrices is called $SL(2,c)$, and $Sp(2)$ is one of
its subgroups.

Wigner's approach is not the only method to obtain the
momentum-preserving transformations.  If we rotate the four-vector
of equation~(\ref{vec11}) around the $y$ axis using the rotation
matrix
\begin{equation}\label{r44}
R(\theta) = \pmatrix{1 & 0 & 0 & 0 \cr
          0 & \cos\theta & 0 & \sin\theta \cr
            0 & 0 & 1 & 0 \cr 0 & -\sin\theta & 0 & \cos\theta} ,
\end{equation}
the resulting four-momentum is
\begin{equation}
\left(E, -p \sin\theta, 0, -p \cos\theta \right)  .
\end{equation}
The above four-by-four matrix corresponds to the two-by-two matrix
$R(\theta)$ of equation~(\ref{r22}).

It is possible to boost this four-vector along the positive $x$
direction using the boost matrix corresponding to $B(2\lambda)$ of
equation~(\ref{barg22}). Then the four-momentum becomes
\begin{equation}
\left(E,  p \sin\theta, 0, - p \cos\theta \right)  .
\end{equation}
The four-by-four boost matrix is
\begin{equation}\label{b02}
B(2\lambda) = \pmatrix{\cosh(2\lambda) & \sinh(2\lambda) & 0 & 0 \cr
           \sinh(2\lambda) & \cosh(2\lambda) & 0 & 0 \cr
            0 & 0 & 1 & 0 \cr 0 & 0 & 0 & 1} .
\end{equation}

We can then rotate this four-momentum to the original form of
equation~(\ref{vec11}), using again the rotation matrix
$R(\theta)$ of equation~(\ref{r44}).  This method was discussed
in detail in Ref.~\cite{hk88}, but only for a massive particle
whose four-momentum takes the form of equation~(\ref{vec11}).

It should be noted that, in order that the four-vector of
equation~(\ref{vec11}) return to the original form after the
rotation-boost-rotation process, the boost parameter $\lambda$
cannot be arbitrary.  It should be determined from the rotation
angle $\theta$ and the four-momentum of equation~(\ref{vec11}).

On the other hand, if the parameters $\theta$ and $\lambda$ are
chosen first, they determine the form of the four-vector.  It
could be time-like (non-zero massive), space-like (for imaginary
mass), or light-like (zero mass), covering all three possible
little groups~\cite{wig39}.

We choose the notation $W$ for the set of the four-by-four matrices
$R(\phi)$ of equation~(\ref{rfi}), $X(\chi)$ of equation~(\ref{xchi}),
and $N(\gamma)$ of equation~(\ref{ngamma}), and call it the
``Wigner matrix.''  Then, in terms of $W$, the matrix $RBR$
can be written as a similarity transformation
\begin{equation}
R(\theta) B(2\lambda) R(\theta) =  S(-\eta) W S(\eta) .
\end{equation}
We are then interested in whether the same relation can be derived
for the two-by-two matrices of the $Sp(2)$ group.

Since the group $Sp(2)$ is locally isomorphic to $O(2,1)$, we
expect to be able to write this similarity transformation in
terms of the corresponding two-by-two matrices.  Indeed, there
are two-by-two $Sp(2)$ matrices corresponding to the four-by-four
matrices for $R(\phi), X(\chi),$ and $N(\gamma)$ given
equation~(\ref{rfi}), equation~(\ref{xchi}), and
equation~(\ref{ngamma}) respectively.  They are
\begin{eqnarray}\label{wm22}
&{}& R(\phi) = \pmatrix{\cos(\phi/2)  & -\sin(\phi/2) \cr
         \sin(\phi/2)  & \cos(\phi/2) } , \nonumber \\[2ex]
&{}& X_{\pm}(\chi) = \pmatrix{\cosh(\chi/2) & \pm \sinh(\chi/2) \cr
      \pm \sinh(\chi/2) & \cosh(\chi/2)} ,\nonumber \\[2ex]
&{}& N_{+}(\gamma) =\pmatrix{1 & 0 \cr \gamma & 1},
\quad  \mbox{or}   \quad
N_{-}(\gamma) = \pmatrix{1 & -\gamma \cr 0 & 1} .
\end{eqnarray}
The two-by-two counterpart of the boost matrix of
equation~(\ref{s44}) is
\begin{equation}\label{s22}
S(\eta) = \pmatrix{e^{\eta/2} & 0 \cr 0 & e^{-\eta/2} } .
\end{equation}
Thus the two-by-two representation of Wigner's little group should
also be written as $S(\mp \eta) W S(\pm \eta)$.  
However, we have to confront the question of which $ N(\gamma)$ 
matrix from equation~(\ref{wm22}) is to be chosen.  
If we go back to the $(RBR)_{\pm}$ matrices of equations~(\ref{rbr}) 
and (\ref{rbrm}),  the upper-right element can vanish for $(RBR)_{+}$,
while $(RBR)_{-}$ can have a vanishing lower-left element.

We shall also see that the transition from $X_{+}(\chi)$ to
$X_{-}(-\chi)$ cannot be achieved through an analytic continuation
from positive $\chi$ to negative $\chi$ in equation~(\ref{wm22}).
We shall assume that $\lambda$ and $\chi$ have the same sign.

Indeed, the two-by-two Wigner matrix has two branches.  Thus
$N_{+}$ and $X_{+}$ should be chosen for $(RBR)_{+}$.  We shall
call this set of Wigner matrices $W_{+}$, and write
\begin{equation}\label{rbr3}
(RBR)_{+} = S(-\eta) W_{+} S(\eta) .
\end{equation}
What then happens to $N_{-}$ and $X_{-}$?

In order to answer this question, let us start with the four-momentum
\begin{equation}\label{vec33}
P = (E, 0, 0, p),
\end{equation}
instead of $(E, 0, 0, - p)$ of equation~(\ref{vec11}).  We can boost
this four-momentum to $E = (m, 0, 0, 0)$, using the inverse of $S(\eta)$
given in equation~(\ref{s44}).  This four-vector is invariant under
rotations.  We can then boost back to the starting form of
equation~(\ref{vec33}).

As for the Bargmann decomposition, we can rotate the four-vector
of equation~(\ref{vec33}) to
\begin{equation}
P = (E, p \sin\theta, 0, p \cos\theta) ,
\end{equation}
by applying the rotation matrix $R(\theta)$.  We then apply the
boost matrix $B(-2\lambda)$ to this four vector to get
\begin{equation}
P = (E, -p \sin\theta, 0, p \cos\theta) .
\end{equation}
The rotation matrix $R(\theta)$ will bring this vector back to
the original form of equation~(\ref{vec33}).  If we complete
this kinematics, the resulting matrix is $(RBR)_{-}$ of
equation~(\ref{rbrm}).  This version of the Lorentz kinematics
was thoroughly discussed in reference~\cite{hk88}.

The resulting $RBR$ matrix is that of equation~(\ref{rbrm}),
which we write as $(RBR)_{-}$.  The lower-left element of
this matrix can vanish.  The corresponding Wigner matrix should
include $N_{-}$ instead of $N_{+}$, and
\begin{equation}\label{rbr4}
(RBR)_{-} = S(\eta) W_{-} S(-\eta) .
\end{equation}
If we combine equation~(\ref{rbr3}) and equation~(\ref{rbr4}),
we can write
\begin{equation}\label{rbr5}
(RBR)_{\pm} = S(\mp \eta) W_{\pm} S(\pm\eta) .
\end{equation}

\section{The Sp(2) matrix as a similarity transformation of the
Wigner matrix}\label{sp2}

Our main concern is to see whether the $M$ matrix or its core $RBR$
of equation~(\ref{rbr}) or equation~(\ref{rbrm}) can be written as a
similarity transformation of the diagonal $D$ matrix of
equation~(\ref{dmat}).  We have already shown that this is not
always possible when the matrix is triangular.  On the other hand
the triangular matrix has one parameter, and is as simple as the
diagonal matrix.

We propose to write this expression as a similarity transformation
of the Wigner matrix $W$ given in equation~(\ref{wm22}).  Let us
combine $RBR$ matrices of equation~(\ref{rbr}) and equation~(\ref{rbrm})
into one expression:
\begin{equation}\label{rbr2}
R(\theta)B(\pm 2\lambda)R(\theta) = \pmatrix{(\cosh\lambda)\cos\theta  &
      -(\cosh\lambda)\sin\theta \pm \sinh\lambda \cr
 (\cosh\lambda)\sin\theta \pm
  \sinh\lambda & (\cosh\lambda)\cos\theta} .
\end{equation}
If the diagonal elements are smaller than one, then
\begin{equation}
(\cosh\lambda)\sin\theta > \sinh\lambda ,
\end {equation}
and the off-diagonal elements of equation~(\ref{rbr2}) have opposite
signs, the $RBR$ matrix can be written as
\begin{equation}
     R(\theta)B(\pm2\lambda)R(\theta) = S(\mp\eta) R(\phi) S(\pm\eta) .
\end{equation}
The right-hand side takes the form
\begin{equation}\label{kalti}
\pmatrix{\cos(\phi/2) & -e^{\mp\eta}\sin(\phi/2) \cr
e^{\pm\eta}\sin(\phi/2) & \cos(\phi/2) }.
\end{equation}
By comparing this expression with equation~(\ref{rbr2}), we
can write $\phi$ and $\eta$ in terms of $\lambda$ and $\theta$,
and the result is
\begin{eqnarray}\label{eta1}
&{}& \cos(\phi/2) = \cosh\lambda \cos\theta,   \nonumber \\[2ex]
&{}& e^{2\eta} = \frac{\cosh\lambda \sin\theta + \sinh\lambda}
     {\cosh\lambda \sin\theta - \sinh\lambda} .
\end{eqnarray}

If the diagonal elements of equation~(\ref{rbr2}) are greater than
one, then
\begin{equation}
(\cosh\lambda)\sin\theta < \sinh\lambda ,
\end {equation}
and the off-diagonal elements have the same sign. We should then
use $X(\chi)$ as the Wigner matrix, and write $RBR$ as
\begin{equation}
R(\theta)B(\pm2\lambda) R(\theta) = S(\pm\eta) X_{\pm}(\chi) S(\mp\eta) .
\end{equation}
The right-hand side of this expression can be written as
\begin{equation}\label{elli}
\pmatrix{\cosh(\chi/2) & \pm e^{\mp\eta}\sinh(\chi/2) \cr
\pm e^{\pm\eta}\sinh(\chi/2) & \cosh(\chi/2) }.
\end{equation}
If we compare this expression with the $(RBR)_{\pm}$ matrix of
equation~(\ref{rbr2}),
\begin{eqnarray}\label{eta2}
&{}& \cosh(\chi/2) = \cosh\lambda \cos\theta,  \nonumber \\[2ex]
&{}& e^{2\eta} = \frac {\cosh\lambda \sin\theta + \sinh\lambda}
  {\sinh\lambda - \cosh\lambda \sin\theta } .
\end{eqnarray}

The upper-right or the lower-left component of equation (\ref{rbr2})
can go through zero starting either from a small negative number or
from a small positive number.   

For the first case, if $(\cosh\lambda \sin\theta - \sinh\lambda)$ goes 
through zero from a small negative to a small positive number,
for the $(RBR)_{+}$ branch one should start from
$S(-\eta)R(\phi)S(\eta).$
Then
\begin{equation}
H_{+}(2\sinh \lambda)=S(-\eta)R(\phi)S(\eta),
\end{equation}
hence in this negative region we have
\begin{equation}\label{nr1}
2\sinh \lambda=e^{\eta}\sin (\phi/2).
\end{equation}
For the  $(RBR)_{-}$ branch one should start from
$S(\eta)X_{-}(\chi)S(-\eta),$
then 
\begin{equation}
H_{-}(-2\sinh \lambda)=S(\eta)X_{-}(\chi)S(-\eta).
\end{equation}
So in this negative region we have
\begin{equation}\label{nr2}
2\sinh \lambda=e^{\eta}\sinh (\chi/2).
\end{equation}

When $\eta$ goes to infinity, the upper-right component
of $S(-\eta)R(\phi)S(\eta)$ section of equation~(\ref{kalti}),
namely $~~-\exp^{-\eta}\sin(\phi/2)~~$ goes to zero.
Similarly, $~~\pm \exp^{-\eta}\sinh(\chi/2)~~$ vanishes, which is 
the lower-left component of $S(\eta) X_{-}(\chi)S(-\eta)$
section of equation equation~(\ref{elli}).
In those cases $\phi$ and $ \chi $ 
should be very small so that the diagonal elements collapse to one.  
This limiting process is known as group contraction, and is 
applicable to various problems in physics~\cite{bk03,kiwi90jm}.

On the other hand when $(\cosh\lambda \sin\theta - \sinh\lambda)$
goes through zero from a small positive number to a small negative
number, on the $(RBR)_{+}$ branch one should start from
$S(-\eta)X_{+}(\chi)S(\eta)$, thus
\begin{equation}
H_{+}(2\sinh \lambda)=S(-\eta)X_{+}(\chi)S(\eta). 
\end{equation}
In this positive region we have the relation between 
the parameters as in equation~(\ref{nr2}).
On the  $(RBR)_{-}$ branch we should start from
$S(\eta)R(\phi)S(-\eta)$ thus
\begin{equation}
H_{-}(-2\sinh \lambda)=S(\eta)R(\phi)S(-\eta).
\end{equation}
So in this positive region we have the relation between 
the parameters as in equation~(\ref{nr1}).

When $(\cosh\lambda\sin\theta - \sinh\lambda) = 0$
the diagonal elements are equal to one, one of the off-diagonal
elements vanishes.
The similarity transformation should be written as
\begin{equation}
H_{\pm}(\sinh\lambda) = S(\mp\eta) N_{\pm}(\gamma)S(\pm\eta).
\end{equation}
Thus
\begin{equation}
 2\sinh\lambda =\gamma e^{\eta} .
\end{equation}
Since there is only one variable for the Bargmann decomposition 
collapses into the Iwasawa decomposition, the variables $\gamma$ 
and $\eta$ become combined into one variable.

Let us now return to the original question of diagonalizing the
$ABCD$ matrix or its Bargmann decomposition given in
equation~(\ref{barg11}).
We now have $RBR$ matrix as a similarity transformation of the $W$
matrix.  The transformation matrix $S$ is given above.  Thus,
the $M$ matrix can be written as a similarity transformation given
in equation~(\ref{barg33}).  The transformation matrix is $G = LS^{-1}$,
where the $L$ matrix is defined in equation~(\ref{r22}).  The $G$
matrix now takes the form
\begin{equation}
G_{\pm} = LS^{\mp} = \pmatrix{e^{\mp\eta/2} \cos(\delta/2) & -
           e^{\pm\eta/2} \sin(\delta/2) \cr
     e^{\mp\eta/2} \sin(\delta/2) &  e^{\pm\eta/2} \cos(\delta/2) } .
\end{equation}

We thus conclude that the $Sp(2)$ matrix cannot always be diagonalized,
but it can be written as a similarity transformation of the Wigner
matrix.   The transformation matrix is given above.

\section{Further Mathematical Details}\label{detail}
The $Sp(2)$ matrices are used in many different branches of physics.
They take different forms in the literature, but they are unitarily
equivalent.  The most convenient way to organize those different
expressions is to recognize that the group $Sp(2)$ is a subgroup of
the six-parameter $SL(2,c)$ group, which is locally isomorphic to
the Lorentz group applicable to three space and one time dimensions.
The three-dimensional rotation group is one of its subgroups.  In
addition, it has three $O(2,1)$ Lorentz groups applicable to two
space and one time dimensions.

The form given in equation~(\ref{barg11}) corresponds to the
$O(2,1)$ group applicable to $z, x$ and $t$, where rotations around
the $y$ axis are allowed.  On the other hand, the expression
\begin{equation}\label{barg44}
\pmatrix{e^{i\theta_1/2} & 0  \cr 0 & e^{-i\theta_1/2}}
\pmatrix{\cosh\lambda & \sinh\lambda
           \cr  \sinh\lambda & \cosh\lambda}
\pmatrix{e^{i\theta_2/2} & 0  \cr 0 & e^{-i\theta_2/2}}
\end{equation}
is quite common in the literature, especially in layer optics.  This
is also the form used by Bargmann in his original paper~\cite{barg47}.
This expression represents the $O(2,1)$ subgroup applicable to
$(x,~y,~t)$ while rotations around the $z$ axis are allowed.  Indeed,
the decomposition of equation~(\ref{barg11}) can be obtained from
equation~(\ref{barg44}) by a conjugate transformation with the
matrix~\cite{gk01}
\begin{equation}
\frac{1}{\sqrt2}\pmatrix{1 & i \cr i & 1} .
\end{equation}

It is also possible to decompose the $Sp(2)$ matrix in terms of the
three matrices of the form
\begin{equation}
\pmatrix{\cos\theta & -\sin\theta \cr \sin\theta & \cos\theta}, \quad
\pmatrix{1 & -\gamma \cr 0 & 1}, \quad \pmatrix{1 & 0 \cr \gamma & 1}.
\end{equation}
These expressions are useful in para-axial lens optics.  They are
discussed in detail in Ref.~\cite{bk01}.

Let us go back to the $ABCD$ matrix of equation~(\ref{abcd}).  In
terms of the Bargmann parameters, they can be written as
\begin{eqnarray}
&{}&  A = - \sinh \lambda \sin \delta +\cosh \lambda \cos \theta \nonumber \\[2ex]
&{}& B = \sinh \lambda \cos \delta - \cosh \lambda \sin \theta \nonumber  \\[2ex]
&{}& C = \sinh \lambda \cos \delta + \cosh \lambda \sin \theta  \nonumber \\[2ex]
&{}& D = \sinh \lambda \sin \delta + \cosh \lambda \cos \theta .
\end{eqnarray}
Conversely, the Bargmann parameters are
\begin{equation}
\tan \theta = \frac{C-B}{A+D}, \qquad \tan \delta = \frac{D-A}{B+C} ,
\end{equation}
and
\begin{equation}
\cosh (2  \lambda) = \frac{1}{2C^{2}}
\left \{(A^{2}+C^{2})(C^{2}+D^{2})-2AD+1  \right \} .
\end{equation}

\section{Physical applications}\label{appl}

The immediate application of this similarity transformation is
in one dimensional crystals requiring repeated applications
of the $Sp(2)$ matrix, such as multilayer optics~\cite{sanch05,gk03},
finite periodic potentials~\cite{sprung93,griff01} and thus surface
problems, and laser cavities~\cite{bk02}.
For those problems, we are confronted with the burden of calculating
$M^N$.  Now, we have
\begin{equation}
M^N = \left(G W G^{-1} \right)^N = G W^N G^{-1},
\end{equation}
where it becomes trivial to calculate $W^N$ for all its three cases.

These days, the diagonalization becomes the central issue when we
deal with ``entanglement'' problems in physics, especially when the
system cannot be diagonalized through a rotation.  The system of two
entangled harmonic oscillators is a case in point~\cite{giedke03,kn05}.
Let us start with the ground-state wave function
\begin{equation}\label{wf01}
\psi_{0}(x_{1},x_{2}) = \frac{1}{\sqrt{\pi}}
\exp{\left\{-{1\over 2}(x^{2}_{1} + x^{2}_{2}) \right\}} .
\end{equation}
Here, the variables $x_1$ and $x_2$ are separable.
When they are coupled, the wave function takes the form
\begin{equation}\label{wf02}
\psi_{\eta}(x_{1},x_{2}) = {1 \over \sqrt{\pi}}
\exp\left\{-{1\over 4}\left[e^{-\eta}(x_{1} + x_{2})^{2} +
e^{\eta}(x_{1} - x_{2})^{2} \right] \right\} .
\end{equation}
This wave function can be expanded as~\cite{kn05,kno79ajp},
\begin{equation}\label{expan1}
\psi_{\eta }(x_{1},x_{2}) = \frac{1}{\cosh(\eta/2)}\sum^{}_{k}
\left(\tanh\frac{1}{\eta}\right)^{k} \phi_{k}(x_{1}) \phi_{k}(x_{2}) ,
\end{equation}
where $\phi_{k}(x)$ is the normalized harmonic oscillator wave
function for the $k-th$ excited state.

This expansion serves as the mathematical basis for squeezed states
of light in quantum optics~\cite{knp91}, as an illustrative example
of Feynman's rest of the universe~\cite{fey72,hkn99ajp}, and more
recently as an entangled oscillator state~\cite{giedke03,kn05}.
In order to obtain equation~(\ref{wf02}) from equation~(\ref{wf01}),
we first have to rotate the coordinate system with the matrix
\begin{equation}
\frac{1}{\sqrt{2}}\pmatrix{1 & -1 \cr 1 & 1} .
\end{equation}
Then we have to squeeze the new coordinates using the matrix
\begin{equation}
\pmatrix{e^{\eta/2} & 0 \cr 0 & e^{-\eta/2}} ,
\end{equation}
which is the same as the matrix given in equation~(\ref{s22}).
Indeed, this system was coupled by q squeeze $S(\eta)$ followed by
a rotation by $45^o$ degrees.

In addition, this paper examines the question of transition from
$R(\phi)$ to $X(\chi)$ via $N(\gamma)$ given equation~(\ref{wm22}).
This is known as the stability problem in cavity physics~\cite{bk02}
and focal condition in lens optics~\cite{bk03}.  While its
mathematical aspect was studied in detail in multilayer
optics~\cite{gk03}, its physical implication is yet to be
studied.  There could be many other transitional problems
like this in physics.

Let us note again that the group $Sp(2)$ is locally isomorphic to
the Lorentz group applicable to two space dimensions and one time
variable.  The transition from $R(\phi)$ to $N(\gamma)$ is known as
a group contraction, which is similar to the transition of the
internal space-time symmetries of massive particles to that of
massless particles in the limit $v \rightarrow c $~\cite{kiwi90jm}.
It is interesting to note that this process can be achieved
analytically in terms of the Bargmann parameters.  This requires
further investigation.

\section*{Concluding Remarks}

Two-by-two matrices appear in almost all branches of physics.
For matrices, the basic question is whether they can be brought
to a diagonal form.  We have addressed this basic question in
this paper.  We started with the most general form of the
$Sp(2)$ matrix and its Bargmann decomposition.  We then showed
that it can also be written as a similarity transformation of
the Wigner matrix, which can be either diagonalized or can be
written as a two-by-two matrix with unit diagonal elements and
one vanishing off-diagonal element.

Also in this paper, we have found that the $Sp(2)$ matrix can
be written both in the Bargmann form of writing it as a product
of three-one parameter matrices and also in the Wigner form
as a similarity transformation of the Wigner matrix.  The fact
that $Sp(2)$ matrix can be written in these two different forms
may lead to further interesting results.

\end{document}